# The First Four Ground-Level Enhancements in the 1940s: Investigation, Digitisation, and Analysis of Forgotten Data

Hisashi Hayakawa* (1-4), Stepan Poluianov** (5), Sergey Koldobskiy (5), Alexander Mishev (5), Nicholas Larsen (1), Inna Usoskina (5), and Ilya Usoskin (5,1)

(1) Institute for Space-Earth Environmental Research, Nagoya University, Nagoya, Japan

(2) Institute for Advanced Research, Nagoya University, Nagoya, Japan

(3) Space Physics and Operations Division, RAL Space, Science and Technology Facilities Council, Rutherford Appleton Laboratory, Harwell Oxford, Didcot, UK

(4) Astro-Glaciology Laboratory, Riken Nishina Centre, Wako, Japan

(5) Sodankylä Geophysical Observatory and Space Physics and Astronomy Research Unit, University of Oulu, Oulu, Finland

* hisashi@nagoya-u.jp

** stepan.poluianov@oulu.fi

## Abstract

Intense solar eruptions occasionally accelerate solar energetic particles (SEPs) and can trigger ground-level enhancements (GLEs). Among the 77 known GLEs, the first four GLEs, #1 – 4 in the 1940s took place before the advent of the standard neutron monitors and were missing from the International GLE Database. This data gap challenged their quantification. To overcome this difficulty, we systematically gathered, digitised, and quantified contemporaneous cosmic-ray records pertaining to these GLEs. These data allow us to reconstruct the temporal evolution, with the 1 – 15 min resolutions, of these GLEs, and broaden their geographical coverage to a global scale. GLEs #1 and #3 exhibited gradual increases in their rise times, measured at 45 ± 15 and 105 ± 15 min, respectively. In contrast, GLEs #2 and #4 both exhibited abrupt increases of 15 ± 15 min. We also compared integral ionisation increase on the standard ionisation chambers and their local geomagnetic cutoff rigidities Pc to qualitatively compare these GLE's spectral hardness: Our result indicates that their spectra are extremely hard for GLEs #2 and #4 and mildly hard for GLEs #1 and #3. GLE #3 showed the greatest integral ionisations for polar detectors among them.

## 1. Introduction

Solar eruptive energy releases, such as flares and coronal mass ejections, can occasionally accelerate charged particles in corona and interplanetary medium, forming solar energetic particles (SEPs), which can be detected in the vicinity of Earth (1; 2; 3). Occasionally, the flux and energy of SEPs appear sufficiently high to be detected by ground-based cosmic-ray monitoring systems, mostly a network of standard neutron monitors (NMs). These phenomena are known as Ground-level Enhancements and abbreviated as GLEs (4; 5). GLEs correspond to the strongest SEP events that are known from direct observations. They have a potential to elevate radiation exposure and produce technological hazards for spacecraft or aircraft (6; 7; 8; 9). Especially, intense GLEs may affect not only crew, passengers, and avionics on the flight altitude but also safety-critical industries such as nuclear power and autonomous vehicles on the ground level (7). Moreover, analyses on the most intense GLEs may bridge the knowledge gap towards very rare once-in-a-millennium extreme solar particle events (ESPEs) known from cosmogenic isotopes (10; 11; 12).

At the time of writing, 77 GLEs have been registered by the network of the ground-based instruments from their first confirmed detections in February 1942 (13; 14) to the latest detections in the 2020s (15; 16; 17; 18; 19; 20). Most of these GLEs, particularly those following GLE #5 in February 1956, have been extensively recorded using the world-wide network of standard neutron monitors (21; 22; 23; 24; 25; 26; 27) and catalogued in the International GLE Database (IGLED; https://gle.oulu.fi). This comprehensive compilation provides a crucial basis for scientific analyses (25), although additional research could potentially allow us to find some more GLEs or sub-GLEs (SEPs that cause cosmic-ray enhancements to polar high-altitude stations but do not cause cosmic-ray enhancements on the sea-level stations), especially for the period prior to the 1970s (28; 29).

GLE #5, recorded on 23 February 1956, has been regarded as a reference event for the studies of ESPEs, owing to its characteristics of the greatest fluence and the hard spectrum (11; 27; 30; 31). This event has been utilised to model worst-case scenarios for radiation doses and assess potential technological impacts at flight altitudes (7), as well as to analyse atmospheric impacts of SEP events (32). Furthermore, GLE #5 serves as a link between contemporary GLE observations and extreme solar events inferred from cosmogenic isotopes present in tree rings and ice cores (10; 12; 33; 34; 35; 36; 37; 38; 39; 40).

Occasional qualitative assessments have compared GLE #5 with the earliest recorded GLEs of the 1940s (23; 41; 42). However, there exists a notable scarcity of information and data regarding the original datasets associated with these early GLEs, as they occurred before the standard NM network started its operation (21; 43). The IGLED includes data only from GLE #5 in 1956 onward, without covering these early GLEs #1 – 4 that took place before the standard NM network was developed. They are difficult to assess quantitatively owing to the uncertainties such as those in the cross-instrument calibrations between the contemporaneous detectors and standard NMs, atmospheric effects to the contemporaneous detectors, or instrumental setups (25; 27). These GLEs were recorded through observations of such devices as ionisation chambers (hereafter ICs), telescopes made of early Geiger–Müller tubes (hereafter GM tubes), electroscopes, and a prototype NM. These detectors have some limitations, such as broad energy response, lower sensitivity, higher data noise, and limited stability, e.g., drifts, compared to the standard NMs (44). These difficulties might have made these instruments miss some of the early GLEs before 1956. In recent scientific discussions, Miroshnichenko (23) listed these GLEs among the most intense events in the observational history following the 15-min peak intensities and normalisation of Duggal (41). Shea and Smart (45) highlighted their limited data availability, noting that only bi-hourly measurements from Carnegie ICs in Cheltenham and Huancayo (Figure 1) were readily accessible to the scientific community. They also reported the presence of original data traces archived at the US National Geophysical Data Centre, alongside hourly measurements done with Japanese ICs (46) (Figure 2) and a prototype NM situated in Manchester, UK (47). So far, none of their time series had been digitally available and thus suitable for analyses, despite such discussions.

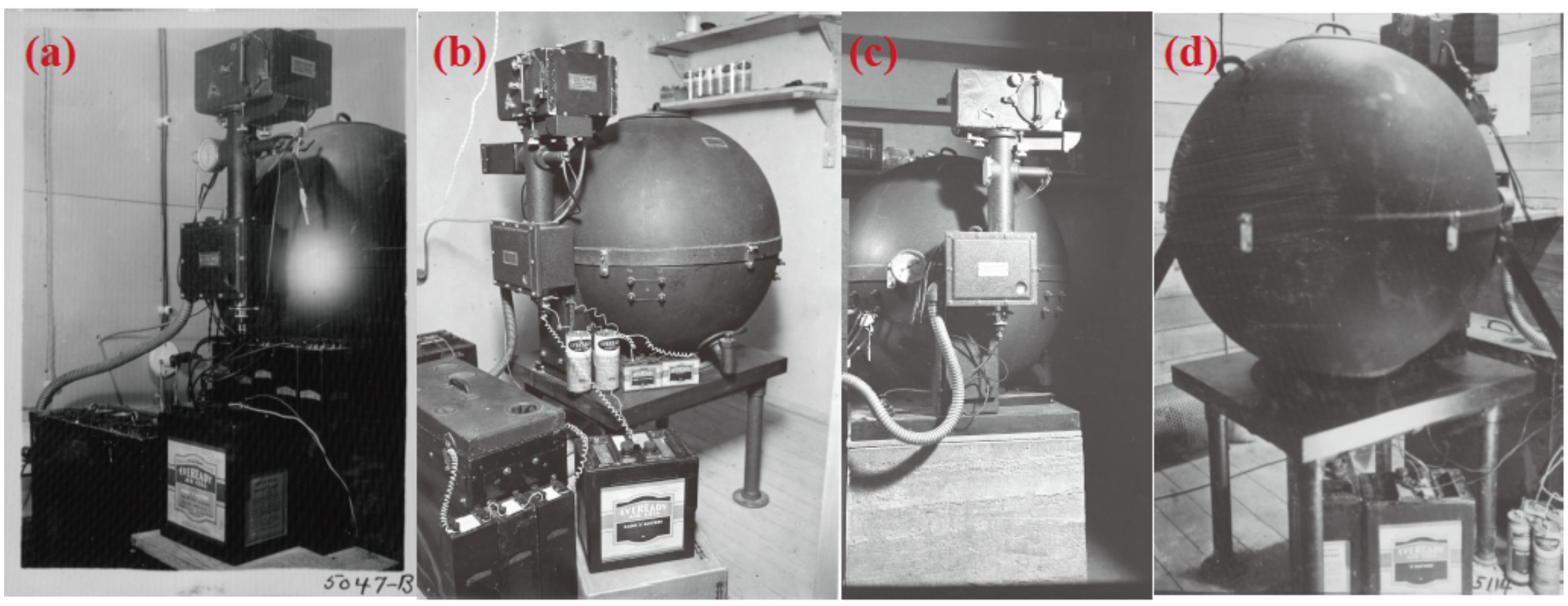

**Figure 1**. Examples of Forbush's ionisation chambers that were operated in (a) Cheltenham

Observatory (5047B), (b) Teoluyucan Observatory (5058), (c) Huancayo Observatory (5080), and (d) Christchurch Observatory (5114), as reproduced from the said archival photographs of Carnegie Science Collections.

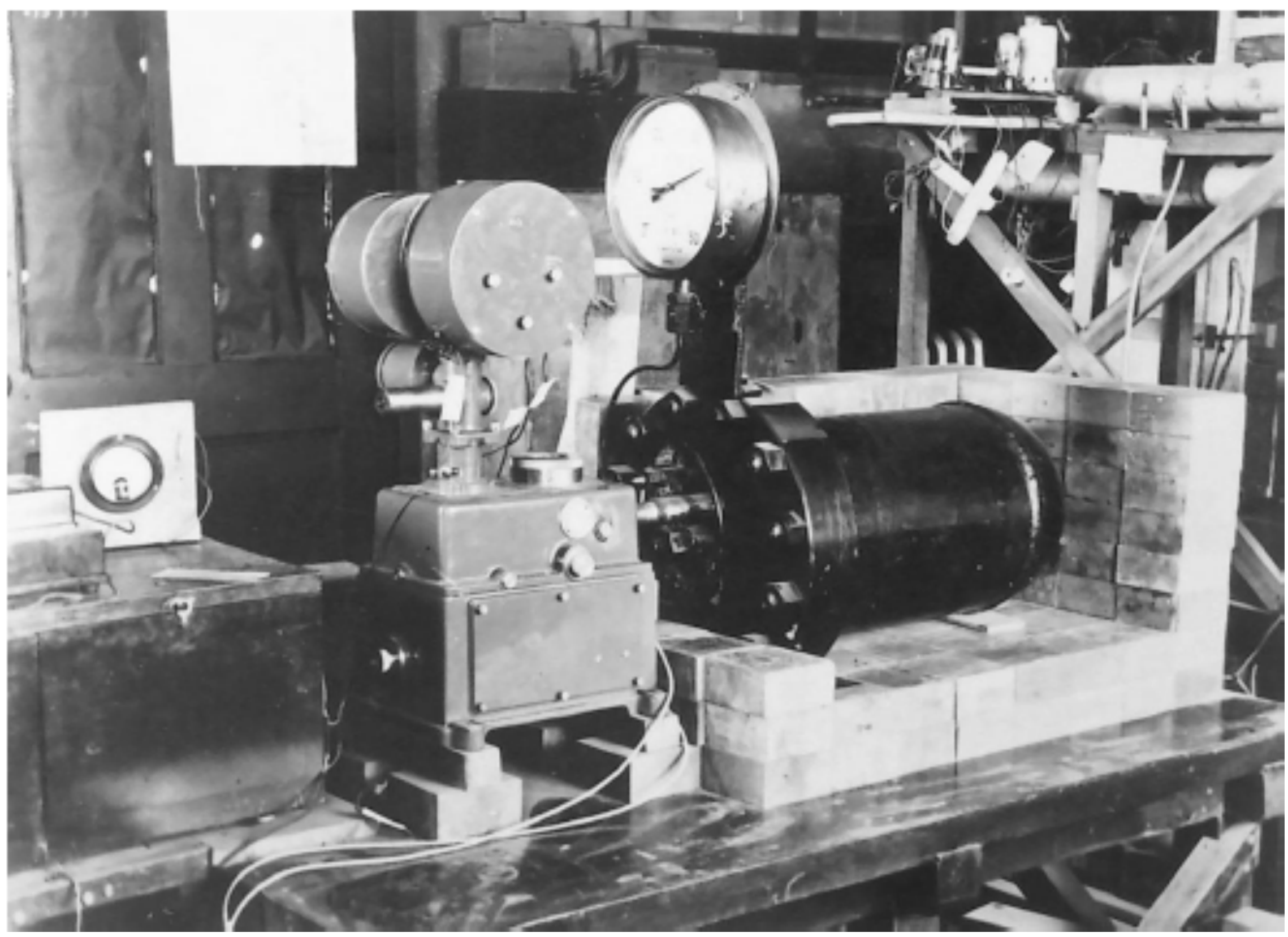

**Figure 2**. One of Nishina-type ionisation chambers that was operated in RIKEN and currently preserved in Yoshio Nishina Memorial Room of RIKEN, as reproduced with courtesy to the Nishina Memorial Foundation.

Importantly, these GLEs were recorded using a variety of contemporary devices (e.g., Figures 1 and 2), with data captured in diagrams and tables that served as contemporaneous scientific reports (48; 49; 50), apart from the bi-hourly datasets of the often-consulted Carnegie IC records. These diagrams potentially provide data at higher temporal resolutions, facilitate discussions on the intricate temporal evolution of GLEs, and enhance geographic coverage. To accommodate that, we comprehensively surveyed contemporaneous reports and digitised the published diagrams and tables related to contemporaneous GLE measurements during the 1940s. Accordingly, we have produced machine-readable datasets for GLEs identified as #1 – 4. These datasets will be integrated into the IGLED in the standard GLE format to cover the complete observational history of GLEs (Sections 2–5). We computed the geomagnetic cutoff rigidity for each station with GLE measurements. We then compared the distributions of the integral percentage increase at each site against their local cutoff rigidity for each of the early GLEs (Section 6). Based on these new datasets, we quantified the GLEs #1 – 4 at the best possible temporal resolutions, in their rise times, integral ionisations, and spectral characteristics, revealing insights that were not discernible in the existing bi-hourly datasets (Sections 7 – 9).

## 2. Materials and Methods

A comprehensive bibliographic survey was first conducted to identify and digitise diagrams that depict the temporal evolution of cosmic-ray measurements published in close temporal proximity to the GLEs #1 – 4. This initial step involved identifying relevant references by examining early reviews (48; 49; 50) and contemporary scientific journals. Following this, diagrams representing contemporaneous cosmic-ray measurements were cropped, and the data points were digitised directly with the tool WebPlotDigitizer, intentionally avoiding the digitisation of curves to prevent the introduction of potentially misleading interpolations. For the majority of data given as relative increases in the units of %, we conservatively estimated the absolute digitisation error as ±(0.1 – 0.3) percentage points varying for different plots due to their quality. Exceptionally, the data for Ottawa, Resolute, and Climax cosmic-ray detectors during GLE #4 have the highest digitisation errors of about 2 percentage points, partially because of their high percentage increase. The data given in absolute units, e.g., count rates or voltage, have the digitisation error of the same order of a per cent. We also conservatively estimated the timing error as ±1 minute for records with a resolution of 5 minutes or less, around ±3 minutes for 15-minute resolution series, and up to ±15 minutes for hourly and bihourly data. Besides, we tried some archival investigations, if not comprehensively. For GLE #4, an unexpected discovery was made at the University of Chicago Archives—a source table related to the work of Adams (47) that John Simpson received from Henry Braddick (Figure 3). This finding emerged as a by-product of the surveys conducted in (27). Therefore, instead of digitising the diagram from (47), this source table was tabulated and compiled into our dataset.

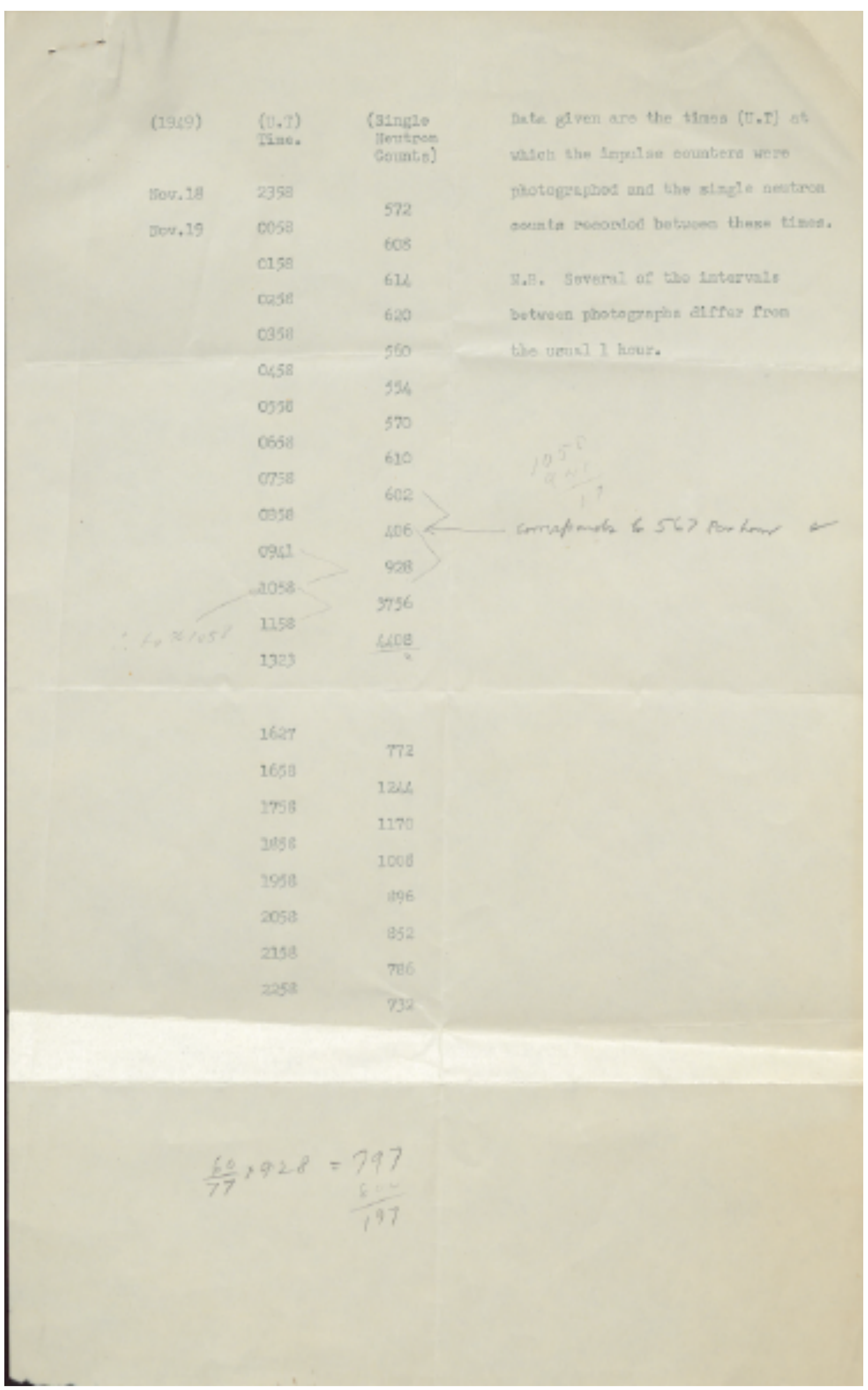

| (1949) | (U.T) Time. | (Single Neutron Counts) |
|---|---|---|
| Nov.18 | 2358 | |
| | | 572 |
| Nov.19 | 0058 | |
| | | 608 |
| | 0158 | |
| | | 614 |
| | 0258 | |
| | | 620 |
| | 0358 | |
| | | 560 |
| | 0458 | |
| | | 554 |
| | 0558 | |
| | | 570 |
| | 0658 | |
| | | 610 |
| | 0758 | |
| | | 602 |
| | 0858 | |
| | | 406 |
| | 0941 | |
| | | 928 |
| | 1058 | |
| | | 3756 |
| | 1158 | |
| | | 4408 |
| | 1323 | |
| | 1627 | |
| | | 772 |
| | 1658 | |
| | | 1244 |
| | 1758 | |
| | | 1170 |
| | 1858 | |
| | | 1008 |
| | 1958 | |
| | | 896 |
| | 2058 | |
| | | 852 |
| | 2158 | |
| | | 786 |
| | 2258 | |
| | | 732 |

Data given are the times (U.T) at which the impulse counters were photographed and the single neutron counts recorded between these times.

N.B. Several of the intervals between photographs differ from the usual 1 hour.

**Figure 3**. The source data table for the Manchester NM in Braddick's correspondence to Simpson in Simpson Folder 216 Box 14, with a courtesy to the Archives of Hanna Holborn Gray Special Collections Research Center, University of Chicago Library

**Table 1**. Cosmic-ray stations used in this work: station name; acronym; type of instrumentation – ionisation chamber (IC), Geiger–Müller tube (GM), neutron monitor (NM), or electroscope (ES); geographical coordinates; altitude above sea level (asl); and data availability for the GLEs (signs "+", "–", and "x" indicate the presence, absence, and textual mention to non-detection, respectively, of records in the dataset). The lower-case "x" in an acronym denotes the number of a specific detector in the location. Altitudes denoted by asterisk were deduced from the station coordinates using OpenTopoMap. We show their values only if the altitude was not given in their source report or in (51).

| Station | Acronym | Instrument | Location | Altitude (m asl) | GLE #1 | GLE #2 | GLE #3 | GLE #4 |
|---|---|---|---|---|---|---|---|---|
| Amsterdam | AMSICx | IC | N52°21′ E4°57′ | 0 | - | + | + | + |
| Azabu | AZABIC | IC | N35°40′ E139°44′ | 27* | x | + | - | - |
| Bargteheide | BARGMx | GM | N53°44′ E10°16′ | 52* | - | - | - | + |
| Cheltenham | CHLTIC | IC | N38°44′ W76°51′ | 72 | + | + | + | + |
| Christchurch | CHRSIC | IC | S43°32′ E172°37′ | 8 | + | + | - | + |
| Climax | CLMXIC | IC | N39°24′ W106°12′ | 3500 | - | - | - | + |
| Freiburg | FREIGM | GM | N47°55′ E7°45′ | 1200 | - | - | - | + |
| Friedrichshafen | FRHNGM | GM | N47°39′ E9°28′ | 403* | + | + | - | - |
| Godhavn | GODHIC | IC | N69°15′ W53°32′ | 9 | + | + | + | + |
| Huancayo | HUANIC | IC | S12°3′ W75°20′ | 3350 | + | + | + | + |
| Komagome | KOMAIC | IC | N35°44′ E139°44′ | 32* | x | + | - | - |
| London | LONDGM | GM | N51°30′ W0°11′ | 24* | + | + | - | - |
| Manchester | MANCGM | GM | N53°28′ W2°14′ | 42* | - | - | + | + |
| Manchester | MANCNM | NM | N53°28′ W2°14′ | 42* | - | - | - | + |
| Moscow | MOSCIC | IC | N 55°29′ E37°19′ | 203 | - | - | - | + |
| Mount Wilson | MTWLES | ES | N34°14′ W118°3′ | 1740 | - | - | + | - |
| Norfolk | NORFIC | IC | N36°52′ W76°18′ | 8* | + | - | - | - |
| Ottawa | OTWGMx | GM | N45°24′ W75°36′ | 101 | - | - | - | + |
| Predigtstuhl | PREDIC | IC | N47°42′ E12°53′ | 1614 | - | - | - | + |
| Resolute | RSLGM3 | GM | N74°41′ W94°55′ | 17 | - | - | - | + |
| Teoloyucan | TEOLIC | IC | N19°45′ W99°11′ | 2285 | + | - | - | - |
| Tokyo | TOKYGM | GM | N35°44′ E139°44′ | 32* | - | - | - | + |
| Thule | TULEES | ES | N76°31′ W68°42′ | 260 | - | - | + | - |
| Weissenau | WEIGMx | GM | N47°46′ E9°36′ | 427 | - | - | - | + |
| Yakutsk | YAKTIC | IC | N62°1′ E129°43′ | 108 | - | - | - | + |

In the summary Table 1, essential details such as the name, geographical coordinates, altitude of each observing station, instrumental category, and presence/absence of the measurement data for each GLE are documented. For acronyms, we use a six-character code, where the first four characters denote the location, and the last two denote the detector type. If multiple detectors of the same type were installed at one location, the final character indicates the detector number, and the part indicating the location is shortened to three characters. The geographical coordinates and altitudes of the individual stations were extracted from their respective source reports. In instances where geographical profiles were not provided, information was sourced from the observatory catalogue maintained by the US Air Force Research Laboratory (51), or alternatively from up-to-date geographical data based on Google Earth Pro. For each location, we computed the geomagnetic cutoff rigidity at each GLE (see Figure 4) using the OTSO tool (52), the International Geomagnetic Reference Field model IGRF 14 (53) for the internal geomagnetosphere, Tsyganenko-89 model (54) for the external geomagnetosphere, and geomagnetic Kp index values corresponding to the time of each GLE peak (55). We have selected the Tsyganenko-89 model over later, more realistic Tsyganenko models, which require spacecraft measurements of the ambient solar wind conditions as input parameters, and such measurements were absent during GLEs # 1 – 4. Tsyganenko-89 model only requires geomagnetic data as the input for a given time; such data is readily available for GLEs # 1 – 4. We note that the Tsyganenko-89 model is found adequate when computing cutoff rigidities under quiet to moderately disturbed geomagnetic conditions (56; 57; 58; 59).

It is unclear whether the source datasets have been routinely corrected for barometric pressure variations, unless explicitly stated in their source documents. In the absence of pressure measurements for each detector, we did not correct our digitisation with barometric pressure in our study. An accurate pressure correction method should be developed in future studies.

If the timestamps were recorded in local time (LT), we corrected the timestamp according to the timezone charts in the reprints of the Admiralty Navigation Manual of the United Kingdom Hydrographic Office (60). Dates of measurements and tables were accessed from https://www.worldtimezone.com.

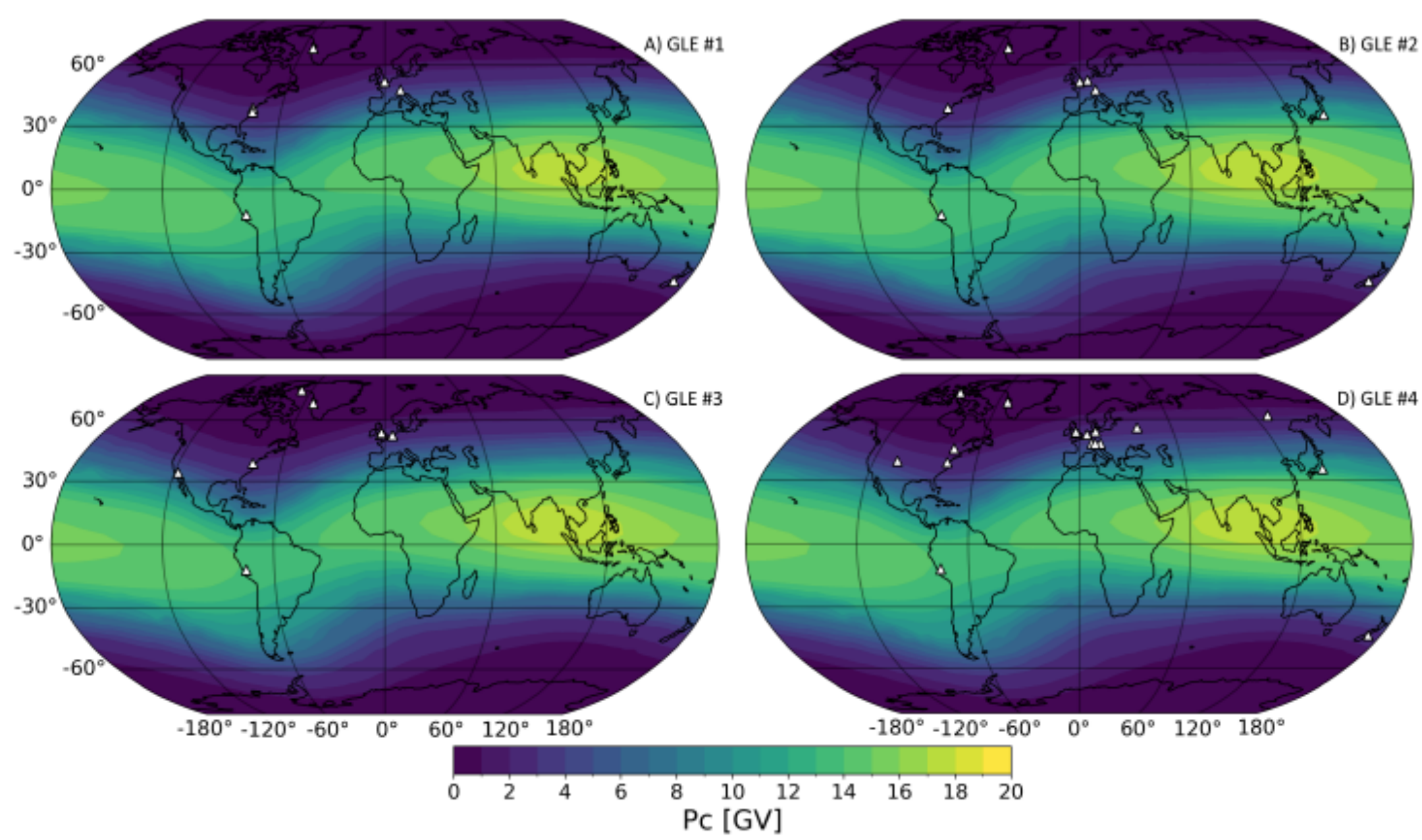

**Figure 4**. World map of geomagnetic cutoff rigidities Pc and stations from Table 1 (triangles) for GLEs #1 – #4 shown in Panels (A)-(D), respectively. The values of Pc are calculated with OTSO (52) specifically for the dates of the GLEs.

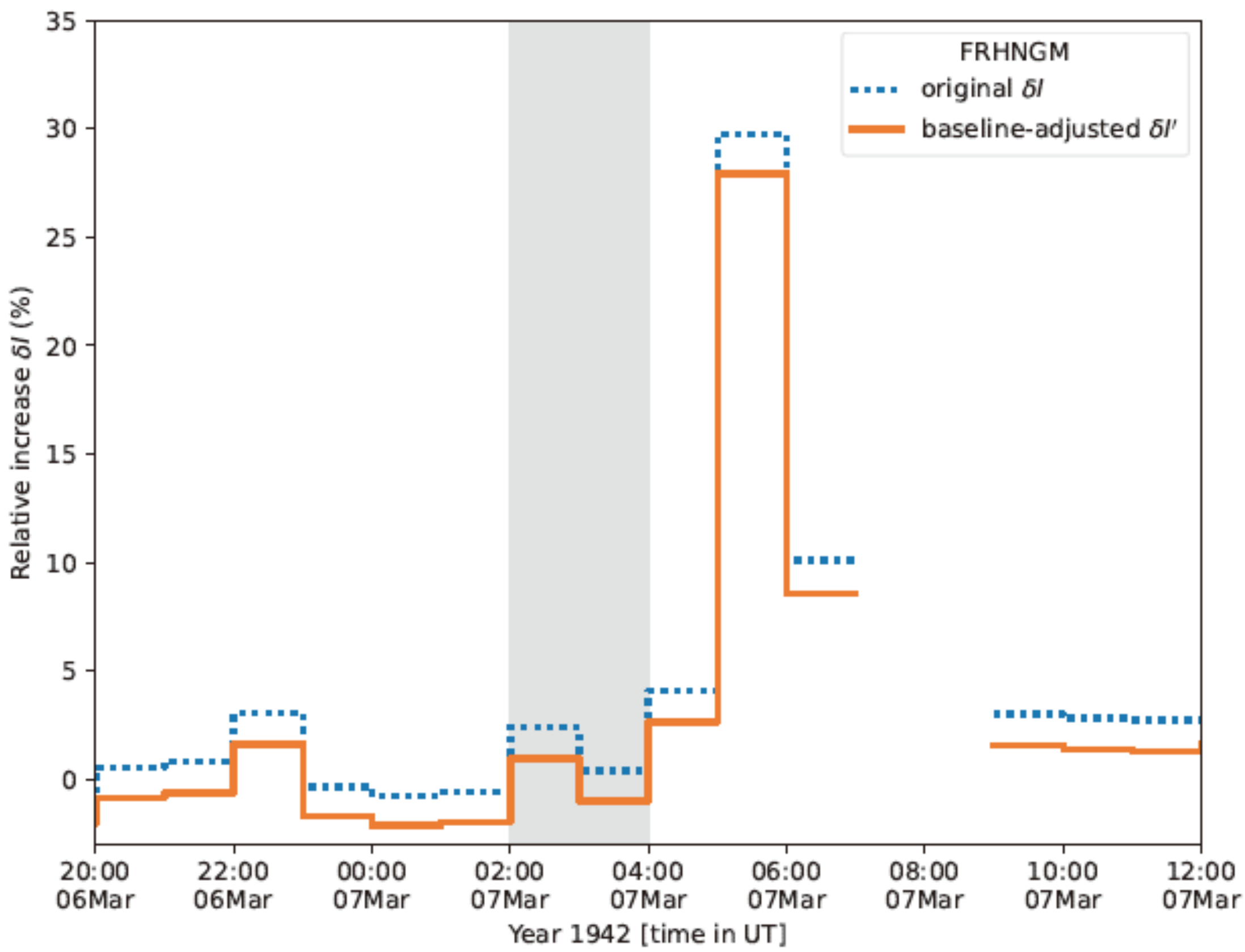

**Figure 5**. Example of the baseline adjustment applied to a relative increase series in this work. Specifically, the original $\delta I$ and adjusted $\delta I'$ records from Friedrichshafen during GLE #2 are plotted as dotted blue and solid orange lines, respectively. The grey area indicates the period 02–04 UT, 7 March 1942, used for calculation of the baseline $\delta I_b$.

Many original data were given as increases $\delta I$ in % relative to baselines defined with different approaches. In order to make the series comparable with each other, we adjusted the recorded data $\delta I$ relative to a standard baseline and computed the standardised relative variation $\delta I'$. In most cases, the original baseline is given as 0% (Figure 5). For its adjustment, we applied the following equation:

$$\delta I'[\%] = \frac{\delta I[\%] - \delta I_{\mathrm{b}}[\%]}{\delta I_{\mathrm{b}}[\%] + 100\%} \cdot 100\%. \qquad (2.1)$$

For the rest of the records, when the original baseline was given as 100% or is absolute units (count rate, voltage, etc.), i.e., without data normalisation, we used the equation

$$\delta I' [\%] = \frac{\delta I - \delta I_\mathrm{b}}{\delta I_\mathrm{b}} \cdot 100\%. \tag{2.2}$$

The standard baseline $\delta I_b$ used in this work is defined based on the average of 2 hours of measurements before the onset of the GLE, following the practice of the database IGLED. Unfortunately, we could not enforce this uniformly for every series, because some of them have missing data points or time resolution that can be low or irregular. In those cases, we tried to use as much data as possible within 2 hours or slightly increase the 2-hour interval for calculating the baseline.

**3. GLEs #1 and #2 of February – March 1942**

GLEs #1 and #2 occurred in a close succession on 28 February and 7 March 1942, with their respective source flares recorded at ≈ 12:00 – 13:00 UT on 28 February (61, p. 318) and 04:14 – 06:20 UT on 7 March (62, p. 202), as indicated by contemporaneous reports of geomagnetic crochets that served as indirect geomagnetic indicators of the solar flares (63). We have collected, digitised, and standardised the contemporaneous measurements in Table 2 and Figure 6 for GLE #1 and Table 3 and Figure 7 for GLE #2, respectively.

**Table 2**. GLE #1 observed by different instruments. The geomagnetic apparent cutoff rigidities Pc are given specifically for the period of that GLE. The resolution corresponds to the interval around the GLE peak. Weak and insignificant peak increases are denoted with a dagger (†) mark. The integral ionisation values are not provided for them. Integral ionisations from data, where there are gaps, are lower estimates and marked with an asterisk (*).

| Station | Reference | $P_c$ (GV) | Peak increase (%) | Best resolution (min) | Integral ionisations (%·h) |
|---|---|---|---|---|---|
| CHLTIC | (13; 14; 64) | 1.63 | 12.75 | 15 | 38.6 |
| CHRSIC | (64) | 2.28 | 7.49 | 15 | 16.7 |
| FRHNGM | (69) | 4.63 | 1.38† | 60 | – |
| GODHIC | (64) | 0.00 | 15.4 | 15 | 39.5 |
| HUANIC | (13; 64) | 14.29 | 0.03† | 120 | – |
| LONDGM | (68) | 2.29 | 1.93 | 120 | 7.2 |
| NORFIC | (65) | 2.02 | 7.44 | 120 | 20.2 |
| TEOLIC | (64) | 10.13 | 0.9† | 120 | – |

**Table 3**. GLE #2 observed by different instruments. The geomagnetic apparent cutoff Pc rigidities are given specifically for the period of that GLE. The resolution corresponds to the interval around the GLE peak. Integral ionisations from data, where there are gaps, are lower estimates and marked with an asterisk (*).

| Station | Reference | $P_c$ (GV) | Peak increase (%) | Best resolution (min) | Integral ionisations (%·h) |
|---|---|---|---|---|---|
| AMSIC5 (12 cm iron shield) | (66) | 2.58 | 8.11 | 60 | 21.0 |
| AMSIC6 (unshielded) | (66) | 2.58 | 3.67 | 120 | 20.8 |
| AMSIC8 (110 cm iron shield) | (66) | 2.58 | 2.86 | 60 | 7.6 |
| AZABIC | (67) | 12.22 | 9.27 | 4 | 4.7 |
| CHLTIC | (13; 14; 64) | 1.92 | 13.78 | 15 | 25.1 |
| CHRSIC | (64) | 2.59 | 12.19 | 15 | 19.4* |
| FRHMGM | (69) | 4.78 | 27.93 | 60 | 39.1* |
| GODHIC | (64) | 0 | 11.97 | 15 | 29.8 |
| HUANIC | (13; 64) | 14.42 | 3.02 | 120 | 5.9 |
| KOMAIC | (46; 67) | 12.22 | 4.89 | 4 | 4.4 |
| LONDGM | (68) | 2.58 | 8.58 | 120 | 25.5 |

During this period, five of Forbush's ICs — Cheltenham, Godhavn, Christchurch, Teoloyucan (the latter operational solely for GLE #1), and Huancayo — were in operation. We digitised their bi-hourly measurements presented in Figure 1 of (64). GLE #1 demonstrated observable spikes in the bi-hourly IC measurements at Cheltenham, Godhavn, and Christchurch; however, no spikes were evident at highcutoff stations of Teoloyucan and Huancayo. In contrast, GLE #2 produced spikes across all bi-hourly IC measurements, including those from Huancayo.

To enhance our analysis, we derived 15-minute interval data for Cheltenham, Christchurch, and Godhavn from Figure 6 of (64), which elucidates the intricate temporal structures associated with these events. GLE #1 began within the 12:00 to 12:15 UT interval at the aforementioned stations, peaking at ≈ 12.8% at Cheltenham during the 13:00 to 13:15 UT interval, ≈ 7.5% at Christchurch at 12:45 to 13:00 UT, and reaching ≈ 15.4% at Godhavn from 13:00 to 13:15 UT. GLE #2 commenced between 05:00 and 05:15 UT at the respective monitoring stations, reaching a peak of ≈ 13.8% at Cheltenham from 05:30 to 05:45 UT. Likewise, Christchurch recorded a peak of ≈ 12.2% from 05:15 to 05:30 UT, while Godhavn observed a peak of ≈ 9.1% within the same 05:15 to 05:30 UT window. Notably, Godhavn experienced a secondary peak reaching ≈ 12.0% between 06:00 and 06:15 UT. These observations delineate the rise time of GLE #2 as 22.5 ± 22.5 minutes.

To create a comprehensive dataset, we integrated the GLE #1 bi-hourly data series from Figure 1 of (64) with the 15-minute data series from the same source (their Figure 6) for Cheltenham, Godhavn, and Christchurch, making necessary adjustments to ensure consistency. Specifically, we validated the 15-minute series, comparing their bi-hourly averages to the published bi-hourly diagrams for their overlaps. A minor discrepancy of less than 1% was detected between the series. We consider the 15-minute series to be more reliable, given the superior quality of Figure 6 in (64) compared to the more generalised overview presented in Figure 1, from which the bi-hourly series was derived.

Additionally, we identified an IC record in Norfolk, United States, as detailed in the work of Berry and Hess (65). We digitised Figure 4 from (65), extracted the hourly data, and adjusted the timestamps from the local time to Coordinated Universal Time (UTC). This process revealed a peak cosmic-ray intensity in Norfolk of ≈ 7.4% occurring around 14:00 UTC during GLE #1.

We also analysed three ICs located in Amsterdam, as documented by Clay et al. (66). These chambers included one unshielded IC (AMSIC6), one shielded with 12 cm of iron (AMSIC5), and another one shielded with 110 cm of iron (AMSIC8). The time resolution for the unshielded IC was bi-hourly, while the shielded ICs operated at an hourly resolution. Clay et al. (66) did not discuss GLE #1, aside from the subsequent Forbush Decrease represented in their Graph 11. Amsterdam ICs might have missed GLE #1. Elliot associated Amsterdam ICs with non detection (48), while it is not clear what source he consulted. In contrast, we successfully identified and digitised their measurements corresponding to GLE #2, which are illustrated in their Graph 12. The recorded peak intensities from these ICs were ≈ 3.7% at 07:00 UTC for the unshielded AMSIC6, ≈ 8.1% at 05:00 UTC for the 12 cm iron-shielded AMSIC6, and ≈ 2.9% at 05:00 UTC for the 110 cm iron-shielded IC AMSIC8 for GLE #2.

Furthermore, in Japan, we located records from one underground IC located in Komagome and three ground-level ICs situated in Azabu (46; 67). These ICs did not observe any significant cosmic-ray increases during GLE #1 (46, p. 3). In contrast, GLE #2 resulted in a substantial increase in cosmic-ray activity across all observed ICs. We digitised their ≈ 5-minute variations from Figure 1 of (67). The cosmic-ray levels recorded at Komagome and Azabu peaked at 4.5% at 05:06 UT and 9.3% at 05:14 UT, respectively. We prioritised the data from (67) over those from (46) because there was a significant contradiction between them, and the former comes from the actual measurers. Unfortunately, presently, it is impossible to resolve the cause of this contradiction, as the original Tokyo IC records were lost at least since 2017, when the Riken shut down the storage at Itabashi.

We verified the functionality of two Geiger–Müller Tube (GM) setups located in the United Kingdom and Germany during this period. In the United Kingdom, we sourced and digitised GM measurements from Imperial College London, as depicted in Figure 3 of (68), enabling us to compile bi-hourly data. This GM effectively detected both GLEs. The peak measurements for GLEs #1 and #2 in London were ≈ 4.0% at 14:00 UT on 28 February and around 8.6% at 06:00 UT on 7 March, respectively.

Meanwhile, in Germany, we gathered and digitised GM readings from Friedrichshafen, illustrated in Figure 1 of (69), allowing us to derive hourly data. This GM also successfully detected both GLEs, with peak values for GLEs #1 and #2 measuring ≈ 1.9% at 13:30 UT on 28 February and an impressive 28.0% at 05:30 UT on 7 March, respectively.

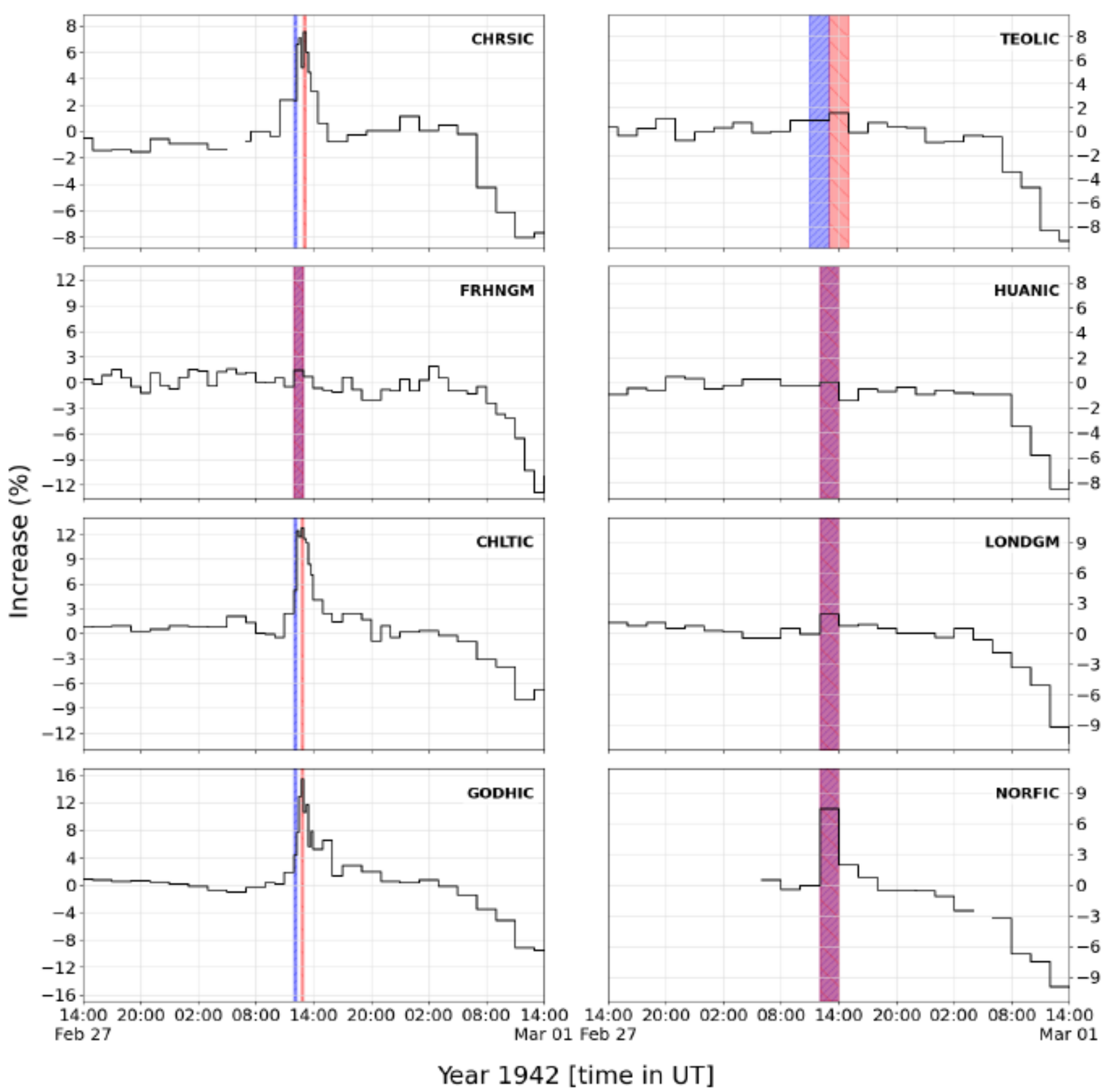

**Figure 6**. Standardised relative increases during GLE #1. The onset period for the GLE and GLE peak in the NM time resolution are denoted by the blue and red shaded regions, respectively. A purple region is used when the onset and peak increase overlap.

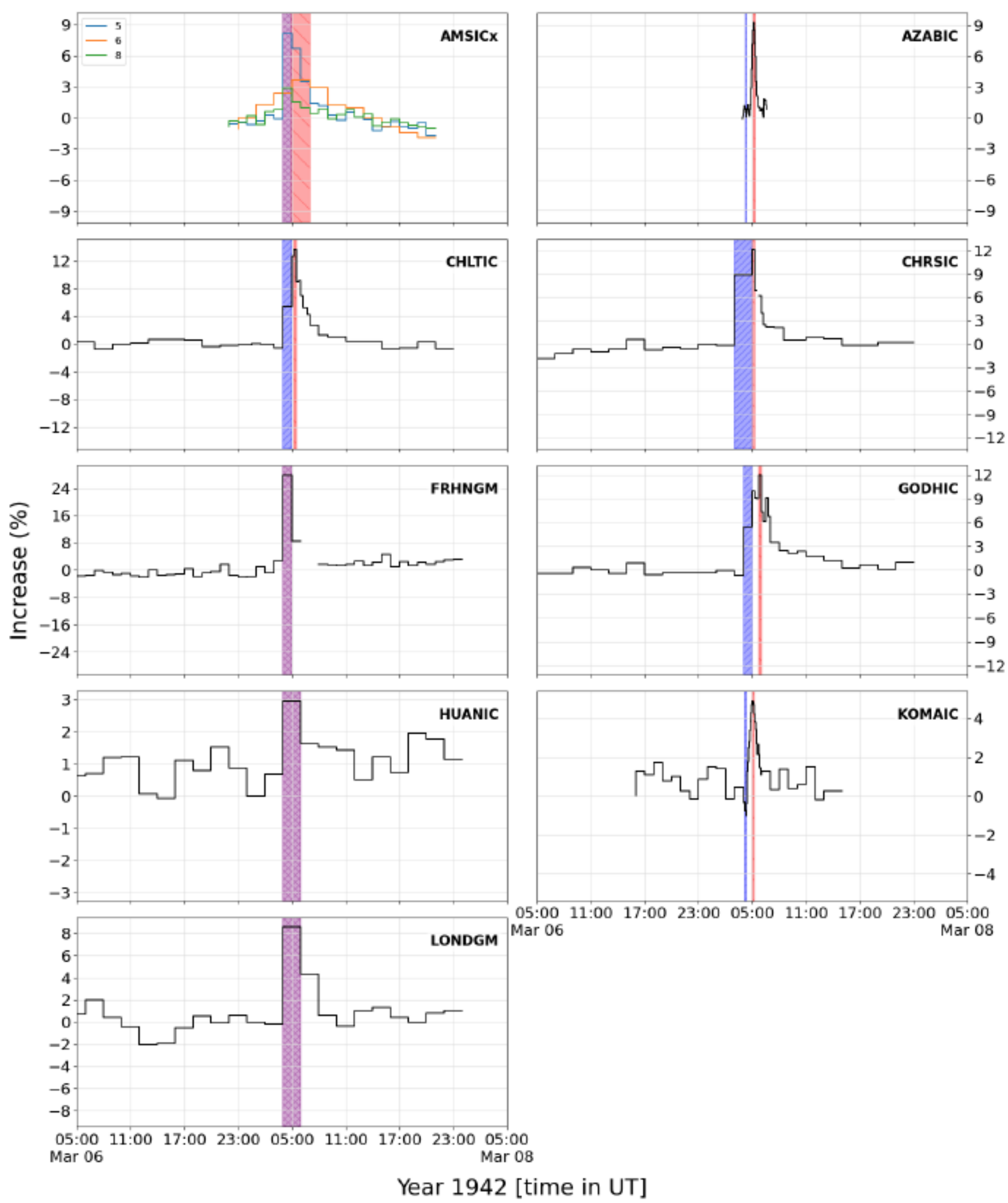

**Figure 7**. Standardised relative increases during GLE #2. Shaded regions for the GLE onset, GLE peak, and any overlapping regions follow the same colour scheme as Figure 6. The GLE peak region for locations with multiple detectors covers the peaks of all detectors at said location.

## 4. GLE #3 of July 1946

GLE #3 was recorded on 25 July 1946. According to contemporaneous Hα flare patrols, the source flare peaked at 16:27 UT on 25 July 1946 (70). This solar eruption triggered an intense and long-lasting GLE. We have collected, digitised, and standardised the contemporaneous measurements in Table 4 and Figure 8 for GLE #3.

At that time, three of Forbush's ICs were operational: those located in Cheltenham, Godhavn, and Huancayo.We have digitised Figure 1 of (71) to include the measurements from these three ICs, which were recorded at hourly intervals around the peak and at bi-hourly intervals thereafter. Notably, measurements at Cheltenham displayed marginal discrepancies when compared to Forbush's bi-hourly dataset for the same observatory, as depicted in Figure 1 of (14). Among the ICs, those at Cheltenham and Godhavn successfully detected GLE #3, whereas the Huancayo IC showed no significant increase during the SEP event.

Furthermore, we have extracted 15-minute interval data for Cheltenham and Godhavn ICs from Figure 7 of (64), which provides a detail of the time structure of their measurements. These measurements reveal notable increases of ≈ 21.8% for Cheltenham and around 16.4% for Godhavn, which were gradually rising for ≈ 120 and 150 minutes for Cheltenham and Godhavn, respectively. Such long rise time is typical for the gradual component of strong GLE events, e.g., GLE #42 (29-Sep-1989) or GLE #77 (11-Nov-2025).

In addition to the Forbush ICs, we also digitised the hourly measurements from two electroscopes that recorded GLE #3: one at Thule in Greenland (72) and the other at the Mt. Wilson Observatory in California (73). The Thule electroscope was protected by 11 cm of lead shielding, whereas the electroscope at Mt. Wilson was unshielded. Specifically, for Mount Wilson, Neher and Roesch (73) reported a 2.2- hour chronological offset between the source flare and the GLE peak. This long duration is not likely instrumental. Their original measurements were taken in the 15-min resolution (73), although they plotted the data only in the hourly resolution.

In England, Dolbear and Elliot (74) operated a GM tube located in Manchester.We digitised their hourly measurements from their figure. At least two additional recorders were operated in Manchester at that time (74), while their data have neither been presented in (74) nor in subsequent related research. The hourly recordings from Manchester indicate a peak associated with the GLE of ≈ 13.3%. Moreover, Dolbear and Elliot (74) synthesised data from three distinct recorders to locate the GLE peak occurring between 18:00 and 18:30 UTC, with a peak value nearing 17% over this half-hour timeframe, although the original source data for this analysis were not explicitly disclosed.

In Amsterdam, four ICs (AMSIC5–8) filled with argon gas at varying pressures of 40, 40, 97, and 58 atmospheres, respectively, were operated. We successfully located and digitised their records, illustrated in Graph 2 of (75). All measurements were presented hourly. Of these, two chambers (AMSIC5 and AMSIC6) were unshielded, while the other two chambers (AMSIC7 and AMSIC8) were covered with 110 cm of iron shielding. We note that the configuration of ICs in Amsterdam during observation of GLE #3 was different from that during GLE #2. These ICs seems to have detected GLE #3, although the open ICs exhibited notable deviations in measurements even before the onset of GLE #3. Each of the four ICs recorded another peak immediately following GLE #3, with the peak magnitudes that varied among the instruments. It remains ambiguous to what extent these ICs were sensitive to external noise. It is not documented what caused this enhancement even before GLE #3. Unless we figure out the actual cause, it is wiser to refrain from using the AMSIC for any future analyses on GLE #3. Therefore, we included the AMSIC dataset neither in Table 4 nor in Figure 8.

**Table 4**. GLE #3 observed by different instruments. The geomagnetic apparent cutoff rigidities Pc are given specifically for the period of that GLE. The resolution corresponds to the interval around the GLE peak. Weak and insignificant peak increases are denoted with a dagger (†) mark. The integral ionisation values are not provided for them.

| Station | Reference | $P_c$ (GV) | Peak increase (%) | Best resolution (min) | Integral ionisations (%·h) |
|---|---|---|---|---|---|
| CHLTIC | (14; 64; 76) | 1.95 | 21.84 | 15 | 110.0 |
| GODHIC | (64; 76) | 0 | 16.44 | 15 | 126.6 |
| HUANIC | (64; 76) | 14.35 | 2.09† | 60 | – |
| MANCGM | (74) | 2 | 13.27 | 60 | 63.5 |
| MTWLES | (73) | 5.54 | 16.37 | 60 | 62.2 |
| TULEES | (72) | 0 | 15.99 | 60 | 200.5 |

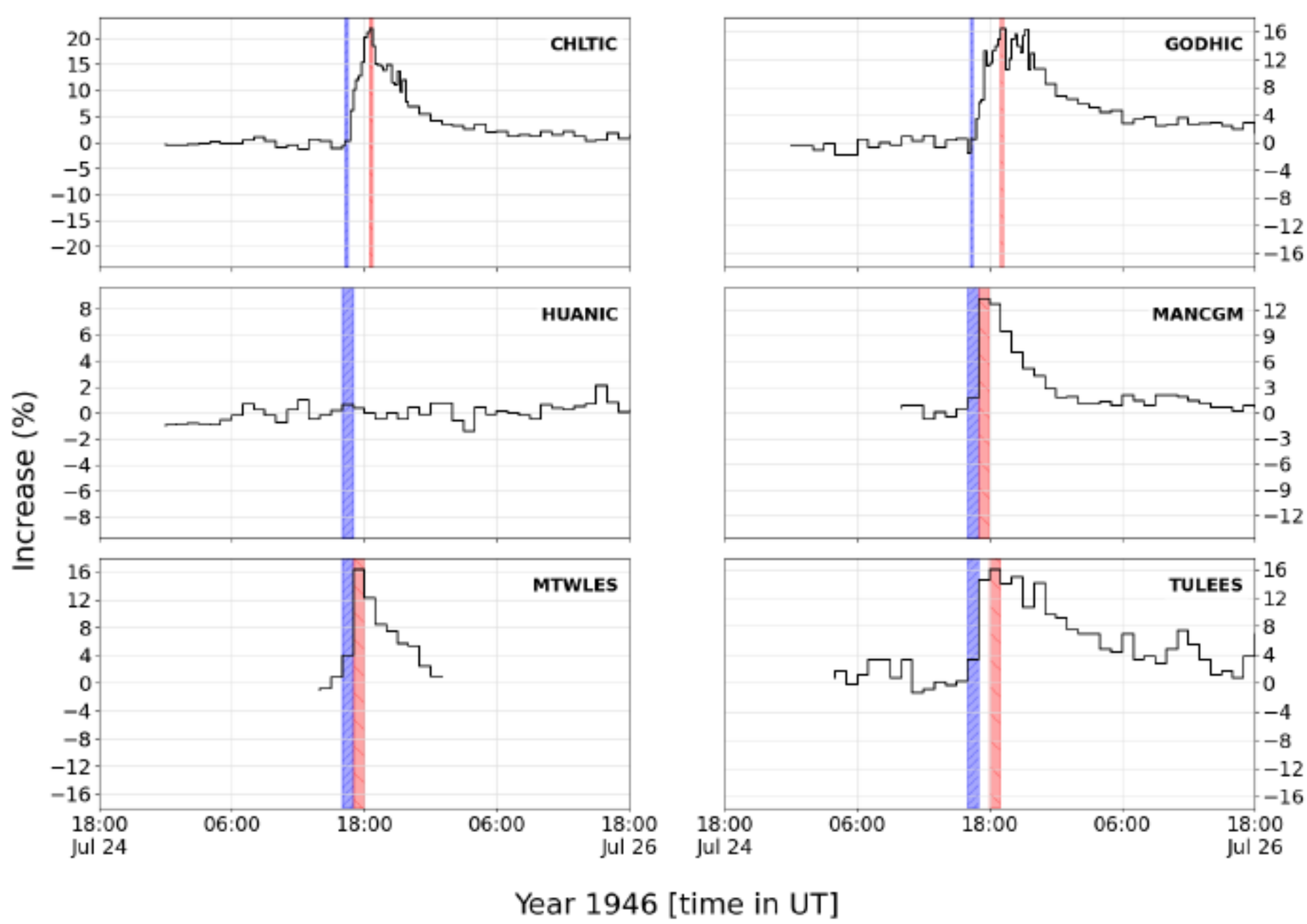

**Figure 8**. Standardised relative increases during GLE #3. Shaded regions for the GLE onset, GLE peak, and any overlapping regions follow the same colour scheme as Figure 6. The GLE peak region for locations with multiple detectors covers the peaks of all detectors at said location. We did not include the AMSICx dataset, which is contaminated by noise of unknown origin and should not be used in further analysis.

**5. GLE #4 of November 1949**

GLE #4 took place on 19 November 1949. According to contemporaneous optical observations (77), the source flare peaked at 10:32 UT on 19 November 1949. We have collected, digitised, and standardised the contemporaneous measurements in Table 5 and Figure 9 for GLE #4.

During this period, five of the Forbush ICs were actively in operation: those located in Cheltenham, Godhavn, Christchurch, Climax, and Huancayo.We digitised the hourly IC measurements from Cheltenham and Huancayo, as depicted in Figure 4 of (64), along with the 15-minute IC measurements from Climax. Furthermore, we digitised Figure 7 of (64), which presents the 15-minute IC measurements for Cheltenham, Christchurch, and Godhavn. Notably, the Huancayo IC recorded a modest GLE spike of ≈ 1.9% between 11:00 and 12:00 UT. In contrast, the Climax IC and the Christchurch IC recorded significant cosmic-ray intensity peaks of around 181% at ≈ 11:00 UT and ≈ 24.2% during the interval from 11:00 to 11:15 UT, respectively. Meanwhile, the Cheltenham and Godhavn ICs got saturated from 11:15–11:45 UT and 11:00– 12:00 UT, respectively, after attaining cosmic-ray intensities of ≈ 40.3% and 26.5%. We showed these intervals as gaps in the respective plots.

This GLE is unique as it marks the inaugural observation of a GLE using a prototype NM (47; 78). We have successfully located and digitised the source table for this measurement at the University of Manchester, alongside correspondence from Henry Braddick, preserved within John Simpson's Box 216 Folder 14, which John Simpson received in 1957 (Figure 3). This documentation allowed us to determine that the peak cosmic-ray intensity registered by the Manchester NM reached ≈ 542.9% during the period from 10:58 to 11:58 UT. The NM was constructed using boron-trifluoride-filled proportional counters, which were integrated into a "Pile" made of high-purity graphite. As reported by Adams (47, p. 504), this configuration demonstrated a detection efficiency for incoming neutrons with energy levels spanning from thermal up to ≈ 10 MeV, with optimal sensitivity concentrated between 3 and 5 MeV. This is because Adams was aware of possible barometric pressure effects on the count rate of the instrument, but the registered variation of the pressure during the reported period of the GLE as within ± 0.1 inches Hg (± 3.4 mb), which would lead to a minor (∓2.5%) variation of the count rate. Moreover, the University of Manchester conducted a separate cosmic-ray measurement using a GM counter, oriented in a north-south direction at a 45◦ angle relative to the local horizon, following the McDonald Elliott configuration (47, p. 503). This measurement documented an increase in cosmic-ray activity of ≈ 12% during the time frame of 11:00 to 12:00 UT (47). We located and digitised a plot for this GM measurement in (49, p. 351) to confirm a cosmic-ray increase of 11.5% at 11:30 UT at a 15-min resolution.

In Japan, we successfully located and digitised cosmic-ray measurements obtained from a counter telescope in Tokyo (identical to a GM detector), as documented by Miyazaki et al. (79). These measurements displayed remarkable temporal resolution. Following the onset of the event at 10:48 UT, the cosmic-ray levels reached a notable peak of ≈ 5.9% between 11:03 and 11:04 UT. This observed value surpasses the peak reported in the running average by Miyazaki et al. (79), which was ≈ 4%. A similar spike in cosmic rays was corroborated by data from a local IC, which was equipped with a 10-cm-thick lead shield beneath a 30-cm-thick concrete roof (79). However, it is regrettable that Miyazaki et al. (79) did not provide any published plots or tables detailing the measurements from the IC.

In the Netherlands, we identified and digitised three IC measurements from Amsterdam, as depicted in Figure 1 of (80). These ICs, filled with argon at a pressure of 60 atmospheres, yielded bi-hourly measurements from one unshielded chamber AMSIC6 and hourly measurements from two shielded chambers – one with a 12 cm iron shield (AMSIC5) and the other with a 110 cm iron shield (AMSIC8). The data indicated that the shielded ICs experienced a peak around 11:00 UT, while the unshielded chamber peaked later, around 12:00 UT. This discrepancy is likely attributable to the less frequent bi-hourly time resolution associated with the unshielded IC. Furthermore, the unshielded AMSIC6 exhibited a significant amount of variability prior to the GLE on 18 November, warranting caution in the interpretation of its data.

In Germany, we located and digitised GM measurements at Predigtstuhl (10 min resolution), Bargtheide with two devices (5 min), and Freiburg (hourly), as depicted in Figures 3, 5, and 6 of (81). A 20- minute GM measurement conducted at Weißenau is illustrated in their Figure 4; however, the data from this measurement reveals a notable gap surrounding the peak. The records indicate that peak cosmic-ray intensity occurred at ≈ 11:40–12:00 UT in Predigtstuhl, 11:55–12:00 UT in Bargtheide I1, 11:45–11:50 UT in Bargtheide I2, and 10:00–11:00 UT in Freiburg.

In France, an IC measurement was recorded at Bagnères (N42◦55′ E0◦11′, altitude 550 m asl) as referenced in (82). This particular instrument was characterised by a large IC fabricated from aluminium– magnesium alloy, having a volume of 120 litres and filled with argon at a pressure of 10 atm. Although Dauvillier's article (82) does not include a graphical representation of the measurement allowing us to digitise their data, it does document a linear rise in cosmic-ray intensity that started around 11:30 UT, reached a maximum of 3.6% at ≈ 12:00 UT, and subsequently decayed until around 15:00 UT.

In the Soviet Union, two IC measurements were performed in Yakutsk and Moscow. The IC measurements from Yakutsk are summarised in Table 1 of (83), where the instrument used was a small S-2 IC, with a volume of 20 litres filled with pure argon. This IC detected a cosmic-ray level peak at approximately 11:30–12:30 UT. We digitised the Moscow IC measurements illustrated in Figure 206 of (49), which revealed a peak occurring between 11:30 and 12:30 UT. This digitisation might be improved if we manage to locate the source IC records.

In Canada, we located five GM measurements from Ottawa (OTWGM1, OTWGM3, OTWGM4, and coincidence counts between detectors GM1–GM2, GM1–GM3) and three from Resolute (RSLGM3 and coincidence counts between GM1–GM2, GM1–GM3), as depicted in Figures 2 and 3 of (84). The captions accompanying these figures indicate that the measurements were conducted at higher time resolutions, specifically from 10:30 to 13:30 UT. Notably, the GM measurements from Ottawa exhibited inconsistent intervals. To enhance the reliability of our data analysis, we recommend substituting these datasets with tabulated original measurements, provided that the source records are retained in Canadian institutions. Although the original data were plotted as relative increases over the background radiation, as summarised in Table 1 of (84), which outlines the average count rates before the GLE, we successfully reconstructed the actual count rates from these relative values. In this work, we decided to omit the coincidence counts data and present only the GM data series. In the records, GLE peaks were observed around 11:00 UT. Conversely, the GM measurements from Resolute employed 20-minute intervals, with peak occurrences noted approximately between 11:30 and 12:30 UT. The peak intensities recorded ranged from 155% to 192% in Ottawa and from 112% to 122% in Resolute, as detailed in Table 1 of (84). This data provides compelling evidence for a significant anisotropic effect during GLE #4.

**Table 5**. GLE #4 observed by different instruments. The geomagnetic apparent cutoff rigidities Pc are given specifically for the period of that GLE. The resolution corresponds to the interval around the GLE peak. Weak and insignificant peak increases are denoted with a dagger (†) mark. Integral ionisations are not provided for them. Integral ionisations from data, where there are gaps, are lower estimates and marked with an asterisk (*).

| Station | Reference | $P_c$ (GV) | Peak increase (%) | Best resolution (min) | Integral ionisations (%·h) |
|---|---|---|---|---|---|
| AMSIC5 (12 cm iron shield) | (80) | 2.64 | 5.64 | 60 | 14.5 |
| AMSIC6 (unshielded) | (80) | 2.64 | 0.59† | 120 | – |
| AMSIC8 (110 cm iron shield) | (80) | 2.64 | 2.94† | 60 | – |
| BARGM1 | (81) | 2.36 | 19.84 | 5 | 26.2 |
| BARGM2 | (81) | 2.36 | 19.91 | 5 | 43.3 |
| CHLTIC | (64) | 1.97 | 40.27 | 15 | 59.6* |
| CHRSIC | (64) | 2.55 | 24.17 | 15 | 43.0* |
| CLMXIC | (64) | 3 | 180.9 | 15 | 301.5* |
| FREIGM | (81) | 4 | 7.57 | 60 | 9.2 |
| GODHIC | (64) | 0 | 26.47 | 15 | 67.74* |
| HUANIC | (64) | 14.31 | 1.88† | 60 | – |
| MANCGM | (49) | 2.1 | 11.5 | 15 | 13.2* |
| MANCNM | (this work, Figure 3) | 2.1 | 542.87 | 60 | 1743.1* |
| MOSCIC | (49) | 2.34 | 8.48 | 60 | 24.8 |
| OTWGM1 | (84) | 0.98 | 69.74 | 2 | 85.7 |
| OTWGM3 | (84) | 0.98 | 69.74 | 2 | 47.8 |
| OTWGM4 | (84) | 0.98 | 69.74 | 2 | 89.6 |
| PREDGM | (81) | 4.25 | 17.88 | 10 | 32.4 |
| RSLGM3 | (84) | 0 | 15.23 | 20 | 39.9 |
| TOKYGM | (79) | 12.26 | 5.88 | 1 | 2.0 |
| WEIGM1 | (81) | 4.13 | 12.16 | 20 | 12.5* |
| WEIGM2 | (81) | 4.13 | 7.75 | 20 | 9.6* |
| YAKTIC | (49; 83) | 1.43 | 14.35 | 60 | 52.2 |

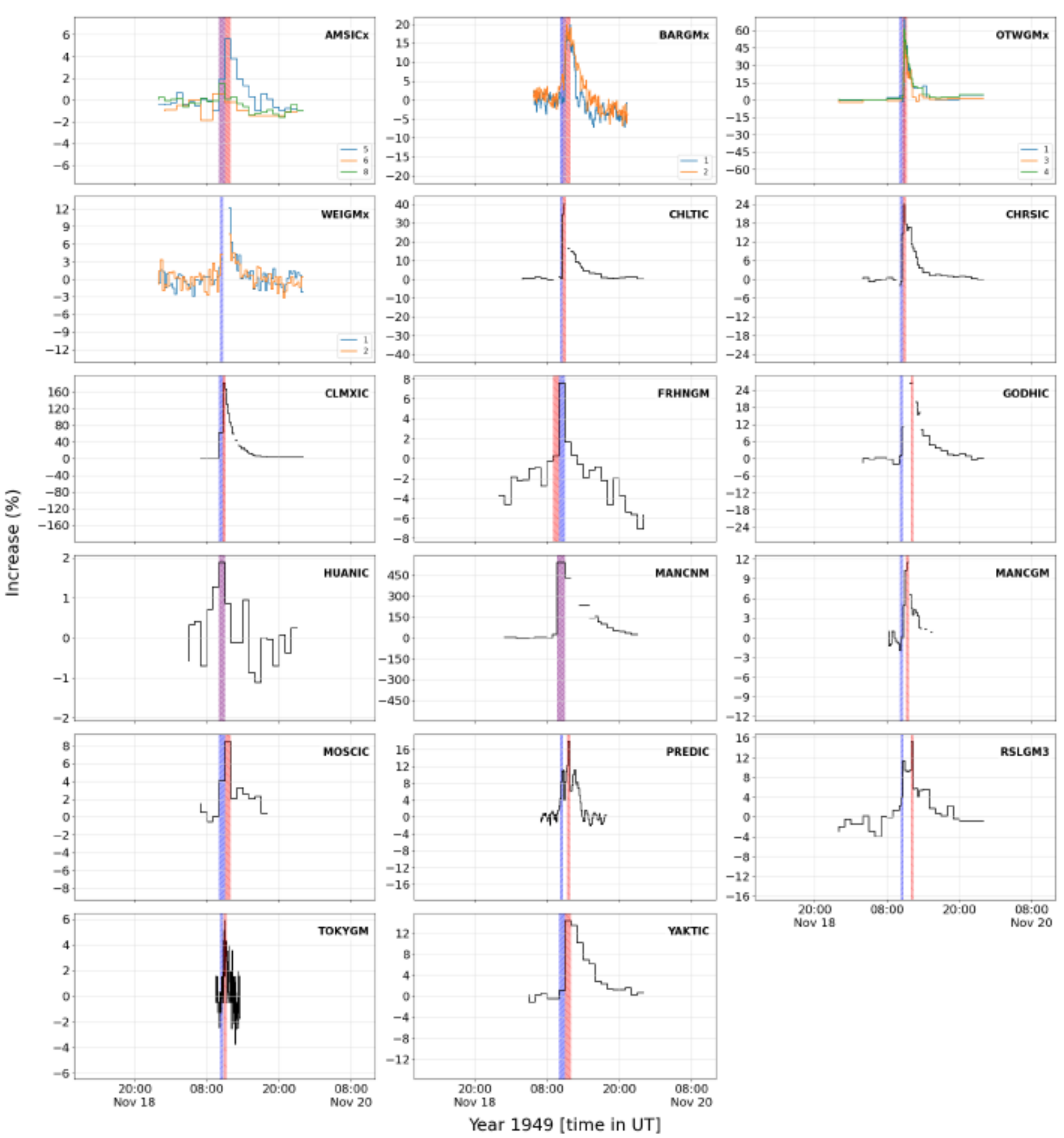

**Figure 9**. Standardised relative increases during GLE #4. Shaded regions for the GLE onset, GLE peak, and any overlapping regions follow the same colour scheme as Figure 6. The GLE peak region for locations with multiple detectors covers the peaks of all detectors at said location.

## 6. Geomagnetic Cutoff Rigidity of the Observational Sites

In Table 1, we present a comprehensive summary of the data recovery based on contemporaneous measurements, and we illustrate their geographical distributions in Figure 4.

Geographical coordinates of these stations (Table 1) allow us to compute an effective geomagnetic cutoff rigidity at each site (Figure 4), using the OTSO package (52) for the geomagnetic conditions during these GLE, the IGRF 14 model (53) for the internal geomagnetosphere, and Tsyganenko-89 model (54) for the external geomagnetosphere. The Tsyganenko-89 model uses the Kp index as an input. The Kp-index values (rounded to the nearest integer) for GLEs #1 – 4 were 6, 2, 4, and 3, respectively (55).

**7. Barometric Corrections**

The majority of the records analysed reflect a percentage increase in relation to the background GCR levels. However, the specific baseline measurements employed are inadequately documented in most of the original reports. The majority of these records did not clarify whether these values were corrected against the barometric pressures. In some cases, it remains ambiguous whether adjustments for barometric pressure were implemented. Contemporary literature offers only partial descriptions, hindering the ability to verify the application of barometric pressure corrections in bi-hourly Carnegie IC records (64, p. 502) and certain Dutch ICs (80, p. 909). This contrasts with the clear absence of such corrections in the Canadian GM measurements (84, p. 232) and the Manchester NM data (47, p. 504). Consequently, further investigations on the original unpublished manuscripts are imperative to determine whether barometric pressure corrections were indeed made to these measurements.

On this basis, it is desired to provide a conservative estimate of the additional uncertainty related to the possible lack of the barometric correction in our dataset. The estimate is based on the fact that a typical barometric sensitivity of a NM is ≈ 0.7% $mb^{-1}$, while that for the ionisation is ≈ 0.4% $mb^{-1}$ for detectors at mid to low latitudes with low altitudes under moderately modulated GCR (85; 86). Considering that the barometric pressure changes would not normally exceed a few mb during the event duration, we conservatively estimate the related uncertainties as 1% and 5%*hr for the peak and integral intensities, respectively. An accidental coincidence of such independent processes as a GLE and a weather storm is unlikely. An explicit account for specific barometric correction of SEPs is not routinely made and is beyond the scope of this work.

**8. Temporal Evolutions**

Our data recovery efforts have facilitated a thorough analysis of the temporal evolution of these GLEs with 15-minute resolutions for the Carnegie ICs. Additionally, we obtained high temporal resolution data, viz.≈ 5 min for GLE #2 and ≈ 1 min for GLE #4. These advancements have significantly contributed to the documentation of the temporal dynamics of early GLEs during the 1940s. For these GLEs, the Carnegie ICs commonly offer measurements at 15-min resolutions, as portrayed in Figures 6 and 7 in (64), although their measurements in Cheltenham and Godhavn were saturated near the peak of GLE #4. The IC measurements were employed to determine the shortest rise times to the first peak for GLEs. Specifically, their shortest rise times can be reconstructed as follows: 45 ± 15 minutes (Christchurch) for GLE #1, 15 ± 15 min (both Christchurch and Climax) for GLE #2, 105 ± 15 min (Cheltenham) for GLE #3, and 15 ± 15 min (Christchurch and Climax) for GLE #4. Our results critically contrast GLEs #2 and #4, both of which were characterised by a prompt rise, with GLEs #1 and #3, which displayed a comparatively gradual rise.

Significantly, for GLEs #2 and #4, we obtained high-time-resolution measurements from Tokyo, Japan, yielding measurement intervals of ≈ 5 min and ≈ 1 min, respectively. These datasets refined measurements allowed us to adjust the shortest rise times for GLE #2 and GLE #4 to 32 ± 4 minutes and 35 ± 1 minute, respectively. These high-time-resolution datasets revealed saturation of the measurements, in contrast to the existing bi-hourly data or the existing 15-min peak values from the Forbush ICs. Certain Carnegie IC measurements exhibited saturation during GLE #4 within the 15-minute dataset, shown as gaps in the respective plots; however, such saturation effects were absent in the bi-hourly data. These saturation issues pose significant challenges in efforts to compare the early GLEs with the contemporary events.

The temporal variations observed in these measurements emphasise the difficulties inherent in quantifying GLEs and accurately determining the peak cosmic-ray increases across different time resolutions. The distinctions among short-rise GLEs become increasingly significant, particularly for GLEs #2 and #4. This pronounced difference is likely attributable to the limited duration of their rise, which proves insufficient for accurate assessments within a bi-hourly time window. Therefore, it is crucial to exercise caution when comparing peak values recorded at different measurement stations, as these stations operate under varying temporal resolutions.

## 9. Integral Ionisation Increases

Expanding the geographical data coverage, our datasets involve richer datasets for detector responses from different Pc ranges (Table 2 – Table 5). On this basis, our results indicate the energy caps of these GLEs on the basis of their highest cutoff rigidities at GLE stations with signal detections, alongside the lowest cutoff rigidities at stations that did not detect GLE signals.

We first examine data from all the detectors to figure out whether they detected the said GLEs or not. We figure out a rigidity cap $P_{cap}$, a boundary for the local detectability for each GLE. For example, when a certain GLE is detected at Tokyo but not detected at Huancayo, the rigidity cap is located between the cutoff rigidity values of Tokyo and Huancayo. According to our datasets, GLE #1 and GLE #3 were not detected in the Huancayo IC. GLE #1 was detected only up to Friedrichshafen station (4.63 GV) but was not detected in Tokyo (11.50 GV). GLE #3 was registered only up to Mount Wilson (5.54 GV) and was likewise not observed in Huancayo, which had a threshold of 14.35 GV. These observations allow us to narrow down the rigidity cap for GLE #1 and GLE #3 to 4.63 – 11.50 GV and 5.54 – 14.35 GV in $P_{cap}$ for the GLE detections, respectively. For the specific case of solar protons, one can use the following equation to covert rigidity $P_{cap}$ to energy:

$$E_{\text{cap}} = \sqrt{P_{\text{cap}}^2 + E_0^2} - E_0 \qquad (9.1)$$

where $E_{cap}$ is the energy in GeV, $P_{cap}$ is the rigidity in GV, and $E_0$ = 0.938 GeV is the rest energy of a proton. Doing so for the found rigidity caps allow us to narrow down the energy caps, an energy threshold for the detections of cosmic-ray precipitations on the ground-based instruments, to 3.8 GeV ≤ GLE #1 ≤ 10.6 GeV and 4.7 GeV ≤ GLE #3 ≤ 13.4 GeV, respectively.

It is highly marginal whether GLE #2 and GLE #4 were detected in Huancayo, while they were robustly detected in Tokyo. Even in the history of the NM measurements, GLEs have been recorded in Huancayo no more than ten times since 1951. ICs or GM counters detected GLEs considerably less frequently than NMs, because the former's energy sensitivity is shifted to higher energies of SEP and GCR with respect to NMs. Contemporary ICs and GM counters also had higher noise levels. Overall, this resulted in the fact that ICs and GM counters registered only the highest-energy and intensity SEP events, much more energetic in comparison to most GLEs registered with NMs.

On this basis, Tokyo detectors at least put an energy cap of about 11.3 GeV ≤ GLE #2 and 11.4 GeV ≤ GLE #4. If we conservatively regard Huancayo data as a non-detection, we can further constrain the range of their rigidity caps as 12.22 GV ≤ GLE #2 ≤ 14.42 GV and 12.26 GV ≤ GLE #4 ≤ 14.35 GV for the GLE detections, respectively. These rigidity caps allow us to narrow down the energy caps as 11.3 GeV ≤ GLE #2 ≤ 13.5 GeV and 11.4 GeV ≤ GLE #4 ≤ 13.4 GeV, respectively. This is more likely the case.

Based on these results, it is desired to qualitatively compare the intensity of these GLEs not merely by the peak ionisation values at each site, but rather through the integral percentage ionisations observed at each site. To this end, we calculated the integral percentage ionisation in cosmic-ray levels relative to the background GCR, based on the data summarised in Table 2 – Table 5 and illustrated in Figure 6 – Figure 9.

It is straightforward to define the duration of each GLE at each station. We set the GLE onset after the flare onset documented in the contemporaneous optical observations or after the onset of the geomagnetic crochets, namely after 12:00 on 28 February 1942 for GLE #1 (61), 04:14 on 7 March 1942 for GLE #2 (62), 16:20 on 25 July 1946 for GLE #3 (70), and 10:30 on 19 November 1949 for GLE #4 (77).

Likewise, we defined the peak time as the time when the count rate reached the highest enhancement in our target period after the GLE onset at each site. We defined the interval between the GLE onset and the GLE peak as a GLE rise time. We have taken the uncertainty range between the lower limits and upper limit, expecting the most extreme scenarios where the GLE onset started in the end of the onset timestamp range and the beginning of the peak timestamp range (a lower limit) and where the GLE onset started in the beginning of the onset timestamp range and the end of the peak timestamp range (a upper limit).

However, it is trickier to quantitatively compare the integrated ionisations at each station for each of these GLEs. This is partly because most of the data that we digitised are from contemporary instruments and require further in-depth analyses for their metadata for the standardisation. For example, the Manchester NMpredated the IGY NMs. It is not immediately clear how to calibrate thisNMwith the IGY NMs. Adams' contemporaneous reports have not told too many details for this issue either (47; 78). Obviously, this NM was much more efficient at detecting cosmic-ray increases in comparison with ionisation chambers. The Manchester NM has an exceptionally high peak value (≈ 542.9%) and would require substantial rescaling when compared to the Manchester GM readings (≈ 11.5%). Each instrument may have different energy sensitivity to cosmic rays. This highlights the necessity for recalibrations in future studies to ensure accuracy and consistency in measurements across different detection systems. Recent archival investigations have revealed substantial amounts of unpublished records for contemporary-device measurements of GLE #5, as a byproduct of our previous survey (27). Their in-depth studies will hint at how to calibrate such early measurements with the modern NM measurements and develop estimates on their absolute magnitudes beyond some preliminary attempts (41; 42). Keeping these caveats in mind, we focused our analyses on measurements from the standardised instruments (Forbush ICs and Nishina ICs; Figures 1 and 2), against their cutoff rigidity for each GLE in Figure 4, to avoid the unwanted uncertainty of the crossinstrumental comparison. These detectors are at least theoretically standardised. In reality, it is known that their measurements drift in a longer time scale (87), while this probably does not affect their short-term responses.

Qualitatively, these data indicate that GLEs #2 and #4 had extremely hard spectra and GLEs #1 and #3 had marginally hard spectra. As we discussed above, GLEs #2 and #4 were detected at least up to Tokyo ICs (12.22 GV for GLE #2 and 12.26 GV for GLE #4). Their detections in Huancayo (14.42 GV for GLE #2 and 14.35 GeV for GLE #4) are at best marginal. Their local integral ionisation should be used as an upper limit for GLEs #2 and #4. Among the Forbush ICs and Nishina ICs, we confirm detections of GLEs #1 and #3 down to Chetltenham and Christchurch, respectively.

These GLEs were all detected in Forbush ICs at Godhavn and Cheltenham. Especially, Godhavn's Pc is calculated as 0 GV for each GLE. While we need to consider shielding of the Godhavn IC in reality, we can, anyway, consider Godhavn as a benchmark for the lowest Pc reference. At Godhavn, the integrated ionisations were 39.5 % h for GLE #1, 29.8 % h for GLE #2, 126.6 % h for GLE #3, > 67.7 % h for GLE #4, respectively. In combination with the energy caps that we discussed above, these data show marginally steep increases for GLE #1 and GLE #3 and gradual increases for GLE #2 and GLE #4. Qualitatively, speaking, they also indicate extremely hard spectra for GLE #2 and GLE #4 and marginally hard spectra for GLE #2 and GLE #4. The integrated ionisations at Godhavn are the most enhanced in GLE #3. This is followed by GLE #4, GLE #1, and GLE #2. This indicates that GLE #3 had the largest fluence among them, GLE #4 had the second largest fluence, and GLEs #1 and 2 were comparable in their fluence, while further intensive studies are needed on how to quantitatively estimate their fluence from integrated ionisations. After all, further analyses are needed to support this simple analysis and to reconstruct their SEP spectra quantitatively.

## 10. Conclusion

In this study, we compiled contemporary cosmic-ray records and measurements related to GLEs from the 1940s. We meticulously organised the source data and digitised the corresponding measurement plots. We constructed time series for 8 stations at GLE #1, 9 stations at GLE #2, 7 stations at GLE #3, and 17 stations at GLE #4. These datasets significantly enhance global data coverage. Moreover, we managed to derive the datasets for the contemporaneous local measurements of the early GLEs in the resolutions of 1 – 15 mins. This improvement allows us to resolve the detailed time structure and integral percentage ionisations of these early GLEs, in contrast with the previous studies that used bi-hourly data from the Forbush ICs of Cheltenham IC and Huancayo (45; 88) and in the existing 15-min peak values from the Forbush ICs (23). These datasets considerably improve the temporal resolution of these events, offering data at 15-minute intervals for GLEs #1 and #3, 5-minute intervals for GLE #2, and 1-minute intervals for GLE #4. This increased resolution has enabled us to ascertain the shortest rise times recorded for these GLEs. Specifically, our findings indicate that GLEs #1 and #3 exhibit gradual rise times of 45 ± 15 minutes and 105 ± 15 minutes, respectively, in contrast to the abrupt rise times of GLEs #2 and #4, both of which occurred within 15 ± 15 minutes.

The expanded geographical coverage allows for a more comprehensive analysis of the GLE spectra by leveraging the terrestrial magnetic field as a natural spectrometer. Our data allow us to qualitatively compare their integral ionisations and spectra with one another, showing hard spectra for GLE #2 and GLE #4, and somehow softer spectra for GLE #1 and GLE #3. Among them, GLE #3 had the greatest enhancements for the integral ionisations.

Our data tables refine the energy caps of these GLEs on the basis of their highest cutoff rigidities at GLE stations with signal detections, alongside the lowest cutoff rigidities at stations that did not detect GLE signals. Notably, GLE #1 was observed in Friedrichshafen, whereas it was not detected in Tokyo. This locates the energy cap of GLE #1 as 3.8 GeV ≤ GLE #1 ≤ 10.6 GeV. GLE #2 was observed in Tokyo but was not detected in Huancayo, indicating its energy cap to be within the range of 11.3 GeV ≤ GLE #2 ≤ 13.5 GeV. Similarly, GLE #3 was observed in Mount Wilson, but not in Huancayo, resulting in locating the energy cap in the range of 4.7 GeV ≤ GLE #3 ≤ 13.4 GeV. GLE #4 was recorded in Tokyo, also absent in Huancayo, locating its energy cap in a range of 11.4

GeV ≤ GLE #4 ≤ 13.4 GeV.

These findings effectively fill the existing data gap for IGLED concerning the initial four GLEs from the 1940s. The compiled data files will enable the quantification of their temporal evolution, the reconstruction of their spectral characteristics and ionisations, and contextualising them with the greatest historical events. Further in-depth analyses are needed on the simultaneous measurements of these non-standard contemporary devices and the standard NMs, especially in the 1950s, to figure out how to calibrate them with the modern datasets.

**Acknowledgements**. HH thanks the University of Chicago for letting us access Braddick's correspondence (Simpson Box 216 Folder 14), Carnegie Science Archives for letting him access and reproduce the historical photos of Carnegie ICs, and Nishina Memorial Foundation for letting him access and reproduce the historical photos of Nishina ICs. HH thanks Kakioka Observatory, Abinger Observatory, Lerwick Observatory, and Eskdalemuir Observatory for providing data for geomagnetic crochets in the 1940s. HH thanks all the cosmic-ray observatories that we mentioned in this article for their operations in the 1940s. HH thanks Jun Nishimura, Yuko Motizuki, and Ryugo Hayano for their help on his search of the Nishina IC measurement records, Kenneth McCracken and Margaret A. Shea for their helpful discussions on the system of the Forbush ICs in the 1940s, Sergey Starodubtsev for providing a copy of (83), and Rachel Hassall, Edmund Henley, and Peter Gallagher for their archival investigations in the United Kingdom and Ireland.

**Funding**. Our research was partially funded by the ISEE International Joint Research Program projects "Digging the past: Uncovering the archival detector data for intense solar particle storms in mid-20 century" and "Solar and terrestrial effects in the 70-year long tritium and Beryllium-10 record from the Dome Fuji in East Antarctica". This research was conducted under the financial support of JSPS Grants-in-Aid JP25K17436 and JP25H00635, the ISEE director's leadership fund for FYs 2021–2025, the Young Leader Cultivation (YLC) programme of Nagoya University, Tokai Pathways to Global Excellence (Nagoya University) of the Strategic Professional Development Program for Young Researchers (MEXT), the young researcher units for the advancement of new and undeveloped fields in Nagoya University Program for Research Enhancement, Transdisciplinary Network linking Space-Earth Environmental Science, History, and Archaeology (JPMXP1324134720) of MEXT Promotion of Development of a Joint Usage/Research System Project: Coalition of Universities for Research Excellence Program (CURE), and the NIHU

Multidisciplinary Collaborative Research Projects NINJAL unit "Rediscovery of Citizen Science Culture in the Regions and Today". This work was partially supported by the ERC Synergy Grant (project 101166910) and the Horizon Europe Program projects ALBATROS (GA 101077071) and SPEARHEAD (GA 101135044).